\documentclass[12pt]{article}

\usepackage{color}
\definecolor{darkblue}{rgb}{0.1,0.1,.7}
\usepackage[colorlinks, linkcolor=darkblue, citecolor=darkblue, urlcolor=darkblue, linktocpage]{hyperref}
\usepackage[]{amsmath}
\usepackage[]{graphicx}
\usepackage[]{latexsym}
\usepackage{slashed,graphicx,color,amsmath,amssymb}
\usepackage[mathscr]{eucal}
\usepackage{mathrsfs}
\usepackage{geometry}
\usepackage[margin=10pt,font=small,labelfont=bf]{caption}
\usepackage{amscd}
\usepackage{bm}
\usepackage{xcolor}
\usepackage{upgreek}
\usepackage[square, comma, sort&compress,numbers]{natbib}
\usepackage[all,cmtip]{xy}
\usepackage[color=cyan!30!white,linecolor=red,textsize=footnotesize]{todonotes}
\numberwithin{equation}{section}
\def\be{\begin{equation}}
\def\ee{\end{equation}}
\def\bea{\begin{eqnarray}}
\def\eea{\end{eqnarray}}
\def\ba{\begin{aligned}}
\def\ea{\end{aligned}}
\def\L{\mathcal{L}}
\def\Q{\mathcal{Q}}

\def\m{\mu}

\def\b{\beta}

 \makeatletter
 \g@addto@macro\bfseries{\boldmath}
 \makeatother

\begin{document}
\vspace*{-.6in} \thispagestyle{empty}
\vspace{.2in} {\large
\begin{center}
{\bf Holographic entanglement entropy in $T\overline{T}$-deformed AdS$_3$}
\end{center}}
\vspace{.1in}
\renewcommand{\thefootnote}{\fnsymbol{footnote}}
\begin{center}
{Miao He$^{a,b}$, Yuan Sun$^c$ 
\vspace{.2in}\\
\textit{$^{a}$School of Physics, Southeast University, Nanjing 211189, China}\\
\textit{$^{b}$Shing-Tung Yau Center, Southeast University, Nanjing 210096, China}\\
\textit{$^{c}$School of Physics and Electronics, Central South University, Changsha 418003, China}\\
E-mail:\ \ \begingroup\ttfamily\small
hemiao@seu.edu.cn,
sunyuan@jlu.edu.cn
\endgroup}
\end{center}

\vspace{.2in}

\begin{abstract}
In this work, we study the holographic entanglement entropy in AdS$_3$ gravity with the certain mixed boundary condition, which turns out to correspond to $T\bar{T}$-deformed 2D CFTs. By employing the Chern-Simons formalism and Wilson line technique, the holographic entanglement entropy in $T\bar{T}$-deformed BTZ black hole is obtained. We also get the same formula by calculating the RT surface. The holographic entanglement entropy agrees with the perturbation result derived from both $T\bar{T}$-deformed CFTs and cutoff AdS$_3$. Moreover, our result shows that the deformed entanglement entropy for large deformation parameter behaves like the entanglement entropy of CFT at zero temperature. We also consider the entanglement entropy of two intervals and study the effect of $T\bar{T}$ deformation on phase transition.
\end{abstract}

\vskip 1cm \hspace{0.7cm}

\newpage

\setcounter{page}{1}
\begingroup
\hypersetup{linkcolor=black}
\tableofcontents
\endgroup
\renewcommand{\thefootnote}{\arabic{footnote}}

\section{Introduction}
\label{sec:1}
The AdS/CFT correspondence gives a geometric interpretation to the conformal field theory. This correspondence allows us to study quantum gravity from the conformal field theory, and it achieves great success in 3D quantum gravity. It is significant to generalize the AdS/CFT correspondence by deforming the conformal field theory and investigating its geometric interpretation. One of the deformed theories called $T\bar{T}$ deformation was proposed and its holographic descriptions were also explored~\cite{Smirnov:2016lqw,Cavaglia:2016oda,McGough:2016lol,Kraus:2018xrn}. It is interesting to establish the holographic dictionary under $T\bar{T}$ deformation. The holographic technique also provides us with a gravitational method to study the $T\bar{T}$ deformed CFT.  
\par
The $T\bar{T}$ deformation is defined through the $T\bar{T}$ flow equation
\begin{align}
\frac{\partial S_{T\bar{T}}}{\partial\mu}=\int d^2x \mathcal{O}_{T\bar{T}},\qquad \mathcal{O}_{T\bar{T}}\equiv T^{ij}T_{ij}+T^2,\nonumber
\end{align}
where $T_{ij}$ is the stress tensor of the deformed theory. This flow equation generates a family of integrable field theory if the original theory is integrable~\cite{Smirnov:2016lqw,Cavaglia:2016oda}. The factorization of $T\bar{T}$ operator leads to the Burgers equation for the deformed spectrum~\cite{Zamolodchikov:2004ce}, so that the spectrum of the deformed theory can be exactly solved. The partition function of the deformed theory can be obtained from various methods, the result turns out that the deformed partition function satisfies a differential equation or an integral transformation of the original one~\cite{Cardy:2018sdv,Dubovsky:2018bmo,Datta:2018thy}. The deformed partition function is still modular invariant~\cite{Aharony:2018bad}. According to the $T\bar{T}$ flow equation, the Lagrangian form and Hamiltonian form were also studied~\cite{Bonelli:2018kik,Jorjadze:2020ili}. There are also some evidences shown that the $T\bar{T}$ deformed theory is a non-local theory~\cite{Dubovsky:2017cnj,Callebaut:2019omt,Tolley:2019nmm,Guica:2020uhm,Guica:2021pzy}. In this irrelevant deformation, it is difficult to study the local properties, such as the correlation function and entanglement entropy. These observables play the important role in the quantum field theory. By using the perturbative method, the correlation functions and entanglement entropy have also been obtained~\cite{Guica:2019vnb,Chen:2018eqk,He:2019vzf,Jeong:2019ylz,Jafari:2019qns,He:2019ahx,He:2020udl,Hirano:2020ppu,He:2020cxp,He:2020qcs,He:2022jyt}. Some non-perturbative results about the correlation function and entanglement were explored in~\cite{Donnelly:2018bef,Chakraborty:2018kpr,Cardy:2019qao,Kruthoff:2020hsi}. However, there is still an open question to calculate the correlation function and entanglement entropy in $T\bar{T}$ deformed theory. For a pedagogical review see~\cite{Jiang:2019epa}.  
\par
According to the AdS/CFT correspondence, the deformed theory can be investigated by using the gravitational approach. There are two points of view to understand the $T\bar{T}$ deformed CFTs from gravity. The one is the $T\bar{T}$ deformed CFTs dual to the AdS$_3$ with a finite radial cutoff~\cite{McGough:2016lol,Kraus:2018xrn}. In this situation, the quasi-local energy of the cutoff region matches the spectrum of the deformed theory. The $T\bar{T}$ flow equation coincides with the Hamilton-Jacobi equation governing the radial evolution of the classical gravity action in AdS$_3$. Many holographic features of the $T\bar{T}$ deformed CFT have been explored based on the cutoff perspective~\cite{Giribet:2017imm,Donnelly:2019pie,Banerjee:2019ewu,Grieninger:2019zts,Caputa:2020lpa,Mazenc:2019cfg,Li:2020pwa,Khoeini-Moghaddam:2020ymm,Li:2020zjb,Kraus:2021cwf}. The other holographic perspective to understand  the $T\bar{T}$ deformation is the AdS$_3$ gravity with certain mixed boundary condition~\cite{Guica:2019nzm}. The boundary condition was derived through the flow equation and variational principle. It turned out that the solution of the metric flow equation related to the higher order Fefferman-Graham expansion, which leads to the mixed boundary condition. For the positive deformation parameter, the mixed boundary condition coincides with the induced metric on the finite radial cutoff. The AdS$_3$ solutions that satisfy the mixed boundary condition were also constructed from AdS$_3$ with Brown-Henneaux boundary condition\cite{Brown:1986nw} through a field-dependent coordinate transformation~\cite{Guica:2019nzm}. The dynamic coordinate transformation approach to $T\bar{T}$ was also found in field theoretic results~\cite{Conti:2018tca,Conti:2019dxg}. The deformed spectrum can also be obtained from the deformed AdS$_3$. Moreover, the mixed boundary condition allows boundary graviton degree of freedom, which turns out to be a $T\bar{T}$ deformed theory~\cite{Ouyang:2020rpq,He:2020hhm,Ebert:2022cle,Kraus:2022mnu}. It also provides us with another approach to studying the $T\bar{T}$ deformation including the holographic entanglement entropy. 
\par
In this paper, we would like to investigate the entanglement entropy in $T\bar{T}$ deformed CFT from holography. For the cutoff perspective, the holographic entanglement was obtained by calculating the length of cutoff geodesic line, and the results match perturbative CFT results~\cite{Chen:2018eqk,Jeong:2019ylz}. The entanglement entropy in $T\bar{T}$ deformation was also studied on both the field theory side and holographic side in recent works~\cite{Asrat:2020uib,Allameh:2021moy,Setare:2022qls,He:2022xkh,Jeong:2022jmp}. We prefer to use the mixed boundary condition perspective to study holographic entanglement entropy. Since the deformed geometry is still AdS$_3$, we will work in the $SL(2,\mathbb{R})\times SL(2,\mathbb{R})$ gauged Chern-Simons formalism of AdS$_3$~\cite{Witten:1988hc}. The Chern-Simons formalism has been used to study $T\bar{T}$ deformation in the literatures~\cite{Llabres:2019jtx,Ouyang:2020rpq,He:2020hhm,He:2021bhj,Ebert:2022cle,Ebert:2022ehb}. In the gauge theory form, the holographic entanglement entropy is encoded in the Wilson line of Chern-Simons~\cite{Ammon:2013hba}, see also~\cite{Huang:2019nfm,Huang:2020tjl}. The Wilson lines are also related to the bulk geometry~\cite{Bakhmatov:2017ihw,Huang:2021qkm}. Generally, the Wilson lines depend on the path and representation of the gauge group. If we choose a appropriate representation of $\mathfrak{sl}(2,\mathbb{R})$, the trace over the representation can be formulated into the path integral of a $SL(2,\mathbb{R})\times SL(2,\mathbb{R})$ invariant auxiliary theory. The on-shell action of the auxiliary is equivalent to the length of geodesics in AdS$_3$. In addition, the Wilson line is a probe in gauge theory, just like a point particle in a curved background. The Wilson lines give a back-reaction to the bulk geometry, and the resulting geometry turns out to be a conical defect on the branch point, which exactly generates a $n$-sheet manifold~\cite{Ammon:2013hba, Huang:2020tjl}. Therefore, the Wilson line back reaction corresponds to the replica trick along the ending points of the Wilson line on the boundary. These results told us that the Wilson line is related to the entanglement entropy through
\begin{align}
S_{\text{EE}}=-\log(W_{\mathcal{R}}(C)), \nonumber
\end{align}
where the ending points of the Wilson line correspond to the interval on the boundary. The thermal entropy also turned out corresponds to the Wilson loop. We use this technique for the deformed AdS$_3$ geometry. The single interval holographic entanglement entropy is calculated exactly, which can reproduce the perturbative result obtained in other literatures~\cite{Chen:2018eqk,Jeong:2019ylz,He:2022xkh}. We also consider the two intervals entanglement entropy in $T\bar{T}$ deformation, which implies a certain phase transition. Moreover, the  holographic entanglement entropy of $T\bar{T}$-deformed AdS$_3$ in the non-perturbative region is also studied. The results show that the entanglement entropy behaves like a zero temperature CFT one for the large deformation parameter.    
\par
The paper is organized as follows: In section~\ref{sec:2}, we give an overview of the gravitational Wilson line approach to obtain the holographic entanglement entropy. In section~\ref{sec:3}, we introduce the deformed AdS$_3$ under $T\bar{T}$, which is parameterized by the deformed spectrum. The holographic entanglement entropy is obtained using the Wilson line approach. We also consider the two intervals entanglement entropy and its phase transition. The same result is derived by calculating the RT surface in the deformed AdS$_3$ in section~\ref{sec:4}. We summarize our results and discussion in section~\ref{sec:5}. The appendix contains our conventions and Wilson line defects.

\section{Wilson lines and entanglement entropy in AdS$_3$}
\label{sec:2}
This section is a review of using the Wilson lines technique to calculate the holographic entanglement entropy, based on~\cite{Ammon:2013hba}. By rewriting the AdS$_3$ gravity in Chern-Simons form, the Wilson line in an infinite-dimensional representation of the bulk gauge group is related to the geodesics in the bulk. According to the Ryu-Takayanagi proposal~\cite{Ryu:2006bv,Ryu:2006ef}, the holographic entanglement entropy or RT surface can be obtained through the Wilson line approach.
\subsection{Wilson lines in AdS$_3$ gravity}
It is well-known that 3D general relativity has no local degrees of freedom, which is purely topological and can be formulated as a Chern-Simons theory~\cite{Witten:1988hc}. In the case of AdS$_3$ gravity, the relevant Chern-Simons gauge group is $SO(2,2)\simeq SL(2,\mathbb{R}) \times SL(2, \mathbb{R})$, so Einstein-Hilbert action can be written as
\begin{align}
S_{\mathrm{EH}}[e, \omega]& = I_{CS}[A]- I_{CS}[\bar{A}],
\end{align}
where the Chern-Simons action is 
\begin{align}
I_{C S}[A]&=\frac{k}{4 \pi} \int_{\mathcal{M}} \text{Tr}\left(A \wedge d A+\frac{2}{3} A \wedge A \wedge A\right),\quad k=\frac{1}{4G}.
\end{align}
The gauge fields $A$ and $\bar{A}$ are valued in $\mathfrak{sl}(2,\mathbb{R})$, which are the linear combination of gravitational vielbein and spin connection
\begin{align}
A=\left(\omega^a+e^a\right)L_a,\quad \bar{A}=\left(\omega^a-e^a\right)L_a.
\end{align}
The $L_a$ are $\mathfrak{sl}(2,\mathbb{R})$ generators, see Appendix~\ref{convention} for our conventions. Variation of the action leads to the equations of motion  
\begin{align} 
F\equiv dA+A\wedge A=0,\quad \bar{F}\equiv d\bar{A}+\bar{A}\wedge\bar{A}=0,
\end{align}
which are equivalent to the first order gravitational field equation and torsion free equation. The AdS$_3$ metric can also be recovered from the gauge fields via
\begin{align}
g_{ij}=\frac{1}{2}\text{Tr}\Big[(A_{i}-\bar{A}_{i})(A_{j}-\bar{A}_{j})\Big].
\end{align}
\par
As a consequence, the AdS$_3$ gravity is formulated into a Chern-Simons gauge theory. By using the Chern-Simons form, we can introduce the gravitational Wilson lines in AdS$_3$ gravity 
\begin{align}
W_{\mathcal{R}}(C) & = \operatorname{Tr}_{\mathcal{R}}\left(\mathcal{P} \exp \int_{C} \mathcal{A}\right),
\end{align}
where $\mathcal{R}$ denotes a representation of $\mathfrak{sl}(2,\mathbb{R})$, and $C$ is a curve on $\mathcal{M}$ with two ending points living on the boundary of $\mathcal{M}$. If the path $C$ is closed, it gives the Wilson loop which is invariant under the gauge transformation
\begin{align}
\mathcal{A}\to \mathcal{A'}=\Lambda^{-1}(d+\mathcal{A})\Lambda.
\end{align}
We can use the Wilson lines to probe the bulk geometry,  instead of a massive particle. The massive particle moving in bulk is characterized by its mass $m$ and spin $s$. These parameters would contribute to the backreaction on the bulk geometry. The trajectory of the particle can be understood as geodesics. When we turn to use the Wilson line to probe the bulk geometry, we have to use the infinite-dimensional representations of $\mathfrak{sl}(2,\mathbb{R})$, characterized by $(h,\bar{h})$. So that the mass $m$ and spin $s$ of the particle can be encoded in the representation of $\mathfrak{sl}(2,\mathbb{R})$ through the relations $m=h+\bar{h}$ and $s=h-\bar{h}$. For the representation of $\mathfrak{sl}(2,\mathbb{R})$ see Appendix~\ref{convention}. 
\par
Note that infinite-dimensional representations of symmetry algebras can be regarded as the Hilbert spaces of quantum mechanical systems in physics. The trace over all the states in the representation $\mathcal{R}$ can be formulated into a path integral of an auxiliary quantum mechanical system. Then the Wilson line can be written as 
\begin{align}
W_{\mathcal{R}}(C) & = \int \mathcal{D} U \exp \left[-S(U ; \mathcal{A})_{C}\right].
\end{align}
where $S(U ; \mathcal{A})_{C}$ is the action of the auxiliary quantum mechanical system that lives on the Wilson line. The action should have a global symmetry group $SL(2,\mathbb{R})\times SL(2,\mathbb{R})$, so that the Hilbert space of the system will be precisely the representation of $\mathfrak{sl}(2,\mathbb{R})$ after quantization.
\par
For the free theory (without gauge fields), an appropriate system is described by a particle moving on the group manifold~\cite{Dzhordzhadze:1994np}, whose action reads
\begin{align}
\label{action-free}
S(U, P)_{\text {free }} & = \int_{C} d s\left(\operatorname{Tr}\left(P U^{-1} \frac{d U}{d s}\right)+\lambda(s)\left(\operatorname{Tr}\left(P^{2}\right)-\mathcal{C}\right)\right),
\end{align}
where $P$ lives in the Lie algebra $\mathfrak{sl}(2,\mathbb{R})$ and $U$ lives in Lie group $SL(2, R)$. The trace in this action means contraction with Cartan-Killing metric. The equations of motion for the free theory are
\begin{align}
\label{eom-00}
U^{-1} \frac{d U}{d s}+2 \lambda P = 0, \\
\label{eom-01}
\frac{d P}{d s} = 0, \\
\label{eom-02}
\text{Tr} P^{2} = \mathcal{C}.
\end{align}
This action has a $SL(2, \mathbb{R})\times SL(2, \mathbb{R})$ global symmetry, namely under the following global gauge transformation
\begin{align}
\label{g-sym}
U(s)\to L U(s)R,\quad P(s)\to R^{-1}P(s)R,\quad L,R\in  SL(2, R),
\end{align}
the action~\eqref{action-free} is invariant. 
\par
In~\cite{Ammon:2013hba}, it turns out that the system coupled with the external gauge fields $A$ and $\bar{A}$ should be 
\begin{align}
\label{couple-gaugefield}
S(U,P;\mathcal{A})_{C} & = \int_{C} ds\left(\operatorname{Tr}\left(P U^{-1} D_{s}U\right)+\lambda(s)\left(\operatorname{Tr}\left(P^{2}\right)-\mathcal{C}\right)\right),
\end{align}
where the covariant derivative is defined by 
\begin{align}
D_{s} U & = \frac{d}{d s} U+A_{s} U-U \bar{A}_{s}, \quad A_{s} = A_{\mu} \frac{d x^{\mu}}{d s}.
\end{align}
The equations of motion become
\begin{align}
\label{eom-1}
U^{-1} D_{s} U+2 \lambda P = 0, \\
\label{eom-2}
\frac{d}{d s} P+\left[\bar{A}_{s}, P\right] = 0,\\
\label{eom-3}
\operatorname{Tr} P^{2} = \mathcal{C}.
\end{align}
After introducing the covariant derivative, the global symmetry \eqref{g-sym} is enhanced to the local gauge symmetry. The action
\eqref{couple-gaugefield} is invariant under local gauge transformation
\begin{align}
\label{l-sym-1}
A_{\mu}&\rightarrow L(x)\left(A_{\mu}+\partial_{\mu}\right) L^{-1}(x), \quad \bar{A}_{\mu} \rightarrow R^{-1}(x)\left(\bar{A}_{\mu}+\partial_{\mu}\right) R(x),\\
\label{l-sym-2}
U(s)&\to L(x^{\mu}(s))U(s)R(x^{\mu}(s)),\quad P(s)\to R(x^{\mu}(s))P(s)R(x^{\mu}(s)).
\end{align}
We have to point out that the equations of motion do not change under these gauge transformations. This feature is useful to construct the solutions of the equations of motion from the free theory solutions. If the gauge fields $A$ and $\bar{A}$ are pure gauge, the solutions for the equations \eqref{eom-1}-\eqref{eom-3} can be obtained from the free theory solution through the gauge transformation \eqref{l-sym-1} and \eqref{l-sym-2}. We will treat more details in section~\ref{sec:HEE-formula}.
\subsection{Equivalence to the geodesic equation}
This Wilson line probe should be equivalent to a massive particle moving
in AdS$_3$. Then we will show that the usual geodesic equation with respect to the metric would appear in the Wilson line path. We denote the Wilson line path in the bulk by $x^{\mu}(s)$. Using the classical equation of motion \eqref{eom-1}-\eqref{eom-3}, the action \eqref{couple-gaugefield} can be reduced into a second order one
\begin{align}
S(U ; A, \bar{A})_{C} & = \sqrt{\mathcal{C}} \int_{C} d s \sqrt{\operatorname{Tr}\left(U^{-1} D_{s} U\right)^{2}}.
\end{align}
In this form, the action is essentially a gauged sigma model, whose equation of motion reads
\begin{align}
\frac{d}{d s}\left(\left(A^{u}-\bar{A}\right)_{\mu} \frac{d x^{\mu}}{d s}\right)+\left[\bar{A}_{\mu}, A_{\nu}^{u}\right] \frac{d x^{\mu}}{d s} \frac{d x^{\nu}}{d s} = 0,
\end{align}
where 
\begin{align}
A^{u}_{s}=U^{-1}\frac{d}{ds}U+U^{-1}A_{s}U.
\end{align}
\par
For the given gauge fields $(A,\bar{A})$, the equation of motion depends on the choice of path $x^{\mu}(s)$. From the perspective of the equation of motion, we learn that $U(s)$ acts like a gauge transformation on the connection $A$. There is a good choice for $U(s)$, so that the particle does not move in the auxiliary space, i.e. $U(s)=\bf{1}$. In this case, the equation of motion reduces to
\begin{align}
\frac{d}{d s}\left(e_{\mu}^{a} \frac{d x^{\mu}}{d s}\right)+\omega_{\mu b}^{a} e_{\nu}^{b} \frac{d x^{\mu}}{d s} \frac{d x^{\nu}}{d s} = 0.
\end{align}
This is precisely the geodesic equation for the curve $x^{\mu}(s)$ on a spacetime with vielbein and spin connection which is equivalent to the more familiar Christoffel symbols forms. Furthermore, the on-shell the action \eqref{couple-gaugefield} for $U(s)=\bf{1}$ becomes
\begin{align}
S(U ; A, \bar{A})_{C} &  = \sqrt{2\mathcal C} \int_{C} d s \sqrt{g_{\mu \nu}(x) \frac{d x^{\mu}}{d s} \frac{d x^{\nu}}{d s}},
\end{align}
which is manifestly the proper distance along the geodesic. 
\par
We have learned that the Wilson line in AdS$_3$ gravity can be expressed as a path integral of an auxiliary quantum mechanical system, whose action is \eqref{couple-gaugefield}. The on-shell action turns out to be the proper distance along the geodesic. Thus in the classical limit, one can find that the value of the Wilson line 
\begin{align}
W_{\mathcal{R}}(x_i,x_f)=\exp(-\sqrt{2\mathcal{C}}L(x_i,x_f)),
\end{align}
where $L(x_i,x_f)$ is the length of the bulk geodesic connecting these two endpoints on the boundary. Holographically, it was proposed by Ryu and Takayanagi that the field-theoretical entanglement entropies correspond to the length of the bulk geodesics ending on the boundary~\cite{Ryu:2006bv,Ryu:2006ef}. In terms of the Chern-Simons description of AdS$_3$ gravity, we can calculate the entanglement entropy from the Wilson line 
\begin{align}
\label{HEE-WL}
S_{\text{EE}}=-\log(W_{\mathcal{R}}(C)).
\end{align}
In~\cite{Ammon:2013hba}, it was also shown that the Wilson line backreaction on the geometry would create a non-trivial holonomy, which can be interpreted as the conical singularity in the bulk. The conical defects hence reproduce the field-theoretical entanglement entropy formula. In the later of this paper, we would like to use the Wilson line technique to compute the holographic entanglement entropy in Chern-Simons AdS$_3$ gravity, including the $T\bar{T}$-deformed AdS$_3$.
\subsection{Holographic entanglement entropy}
\label{sec:HEE-formula}  
In this section, we calculate $W_{\mathcal{R}}(C)$ with $C$ ending on the AdS$_3$ boundary at two points denoted by $x_i=x(s_i),x_f=x(s_f)$. Classically, we just need to calculate the on-shell action of the auxiliary system
\begin{align}
\label{on-shell-action}
S_{\text {on-shell}} & = \int_{C} d s \operatorname{Tr}\left(P U^{-1} D_{s} U\right) = -2 \mathcal{C}\int_{s_{i}}^{s_{f}} d s \lambda(s),
\end{align}
which depends on the solution of the equations of motion. The solutions can be constructed from the free theory solutions, i.e. \eqref{eom-00}-\eqref{eom-02}, through the gauge transformation~\eqref{l-sym-1} and \eqref{l-sym-2}. First of all, we should note the solutions to free theory, denoting them by $U_0(s)$ and $P_0$, are
\begin{align}
U_0(s)=u_0\exp(-2\alpha(s)P_0),\quad \text{with}\quad \frac{d\alpha(s)}{ds}=\lambda(s),
\end{align}
where $P_0$ and $u_0$ are constant. Next, we assume the bulk gauge fields are in pure gauge
\begin{align}
A=L(x)dL^{-1}(x),\quad \bar{A}=R^{-1}(x)dR(x).
\end{align}
In fact, most of the AdS$_3$ solutions are in pure gauge, such as BTZ black hole and Ban\~ados geometry. Then one can verify the following is the classical solution of \eqref{eom-1}-\eqref{eom-3}
\begin{align}
\label{u-p}
U(s)=L(x(s))U_0(s)R(x(s)),\quad P(s)=R^{-1}(x(s))P_0R(x(s)).
\end{align}
These solutions are directly obtained from the local gauge symmetry of the equations of motion.
As argued in~\cite{Ammon:2013hba}, the boundary conditions for $U(s)$ on the boundary ending points can be chosen as 
\begin{align}
\label{ui}
U(s_{i})=&L(x(s_{i}))u_0\exp(-2\alpha(s_{i})P_0)
R(x(s_{i}))=\textbf{1},\\
\label{uf}
U(s_{f})=&L(x(s_{f}))u_0\exp(-2\alpha(s_{f})P_0)
R(x(s_{f}))
=\textbf{1}.
\end{align}
We then have to eliminate the initial value $P_0$ and $u_0$. Solving $u_0$ from \eqref{ui} and substituting into \eqref{uf}, one can find 
\begin{align}
\exp(-2\Delta\alpha P_0)=&R(x(s_i))L(x(s_i))L^{-1}(x(s_f))R^{-1}(x(s_f)).
\end{align}
Taking the trace on both sides, we arrive at 
\begin{align}
\cosh\left(-2\Delta\alpha\sqrt{2\mathcal{C}}\right)=\frac{1}{2}\text{Tr}\Big(R(x(s_i))L(x(s_i))L^{-1}(x(s_f))R^{-1}(x(s_f))\Big),
\end{align}
where we have used 
\begin{align}
\text{Tr}\left(\exp(-2\Delta\alpha P_0)\right)=2\cosh\left(-2\Delta\alpha\sqrt{2\mathcal{C}}\right).
\end{align}
Finally, according to~\eqref{HEE-WL}, we obtain the holographic entanglement entropy formula
\begin{align}
\label{EE-form}
S_{\text{EE}}=&\sqrt{2\mathcal{C}}\cosh^{-1}\left[\frac{1}{2}\text{Tr}\Big(R(x(s_i))L(x(s_i))L^{-1}(x(s_f))R^{-1}(x(s_f))\Big)\right].
\end{align}
We then use this formalism to check the holographic entanglement entropy in Poincare AdS$_3$ and BTZ black hole. 
\subsubsection{Poincar\'e AdS$_3$}
For the case of Poincare AdS$_3$, the line element reads
\begin{align}
ds^2=\frac{dr^2}{r^2}+r^2(d\theta^2-dt^2).
\end{align}
In terms of the Chern-Simons gauge connection, this geometry is described by
\begin{align}
A=&\frac{dr}{r}L_0+rL_{1}(d\theta+dt),\\
\bar{A}=&-\frac{dr}{r}L_0-rL_{-1}(d\theta-dt).
\end{align}
The gauge connections can be written in pure gauge form
\begin{align}
\label{p-adsL}
A=&LdL^{-1},\quad L=\exp(-\ln rL_0)\exp(-(\theta+t) L_1),\\
\label{p-adsR}
\bar{A}=&R^{-1}dR,\quad R=\exp((\theta-t) L_{-1})\exp(-\ln rL_0).
\end{align} 
In order to calculate the entanglement entropy, we consider a time slice ($t=0$) of this geometry and  impose the following boundary conditions for the ending points of the Wilson line
\begin{align}
\label{bdy-condtion-line}
&r(s_i)=r(s_f)=r_0,\\
&\Delta\theta=\theta(s_f)-\theta(s_i)=l,
\end{align}
which means we work on a constant radial boundary and the length of the interval is $l$. Plugging \eqref{p-adsL} and \eqref{p-adsR} into \eqref{EE-form}, one obtains 
\begin{align}
S_{\text{EE}}=&\sqrt{2\mathcal{C}}\cosh^{-1}\left(1+\frac{r_0^2l^2}{2}\right).
\end{align}
Then taking the limit $r_0\gg 1$, so that the result corresponds to the theory living on the conformal boundary, we arrive at~\footnote{We have used the relation
\begin{align}
\cosh^{-1}(x)\sim \log(2x)\quad \text{for}\quad x\gg 1.\nonumber
\end{align}
}
\begin{align}
S_{\text{EE}}&=\frac{c}{3}\log\left(\frac{l}{\epsilon}\right).
\end{align}
where the UV cutoff of the boundary field theory corresponds to the radial cutoff in the bulk, and the central charge relates to the expectation value of Casimir 
\begin{align}
\label{c-epsilon-relation}
\epsilon=\frac{1}{r_0},\quad \sqrt{2\mathcal{C}}=\frac{c}{6}.
\end{align}
The relation between the expectation value of Casimir and central charge can be derived by calculating the Wilson line defect, for the details see Appendix~\ref{app:defect}. This result is exactly the entanglement entropy of CFT$_2$. The same answer can also be obtained by solving the bulk geodesic equation. However, in terms of the Wilson line form, we do not require the solution of any differential equations and follow from purely algebraic operations. This technique can be used for more complicated AdS$_3$ geometry.
\subsubsection{BTZ black hole}
For the BTZ black hole, the metric in Fefferman–Graham gauge is
\begin{align}\label{BTZme}
ds^2=\frac{dr^2}{r^2}+r^2\left(dzd\bar z+\frac{1}{r^2}\mathcal{L}_0dz^2+\frac{1}{r^2}\mathcal{\bar L}_0d\bar z^2+\frac{1}{r^4}\mathcal{L}_0\mathcal{\bar L}_0dzd\bar z\right),
\end{align}
where $\mathcal{L}_0$ and $\mathcal{\bar L}_0$ are constants related to the mass and angular momentum of the black hole
\begin{align}\label{BTZL0}
\mathcal{L}_0=\frac{M-J}{2},\quad \mathcal{\bar{L}}_0=\frac{M+J}{2}.
\end{align}
The corresponding Chern-Simons gauge connections read
\begin{align}
A=&\frac{dr}{r}L_0+\left(rL_{1}-\frac{1}{r}\mathcal{L}_0L_{-1}\right)dz,\\
\bar{A}=&-\frac{dr}{r}L_0+\left(\frac{1}{r}\mathcal{\bar L}_0L_{1}-rL_{-1}\right)d\bar z.
\end{align}
In this case, one can obtain
\begin{align}
L\left(r,z,\bar{z}\right) & = \exp \left(-\ln r L_{0}\right) \exp \left(-zL_{1}+\mathcal{L}_0zL_{-1}\right),\\
R\left(r,z,\bar{z}\right) & = \exp \left(\mathcal{\bar{L}}_0\bar{z}L_{1}-\bar{z}L_{-1}\right) \exp \left(-\ln r L_{0}\right).
\end{align}
In addition, such solutions can be parametrized as
\begin{align}
\label{radial-gauge}
A=b^{-1}(d+a)b,\quad \bar{A}=b(d+\bar{a})b^{-1},\quad b=e^{\ln r L_0},
\end{align} 
Then $a,\bar{a}$ are also flat connections, but do not depend on the radial coordinate
\begin{align}
a=&\left(L_{1}-\mathcal{L}_0L_{-1}\right)dz,\\
\bar{a}=&\left(\mathcal{\bar L}_0L_{1}-L_{-1}\right)d\bar z.
\end{align}
\par
Following the same steps in pure AdS$_3$ and the boundary conditions for the ending points of the Wilson line, we can get
\begin{align}
&\text{Tr}\Big(R(r_0,\theta(s_i),0)L(r_0,\theta(s_i),0)L^{-1}(r_0,\theta(s_f),0)R^{-1}(r_0,\theta(s_f),0)\Big)\nonumber\\
=&-2 \cosh \left(l \sqrt{\mathcal{L}_0}\right) \cosh \left(l\sqrt{\mathcal{\bar{L}}_0}\right)+\frac{\left(\mathcal{L}_0 \mathcal{\bar{L}}_0+r_0^4\right) \sinh \left(l\sqrt{\mathcal{L}_0}\right) \sinh \left(l\sqrt{\mathcal{\bar{L}}_0}\right)}{r_0^2\sqrt{\mathcal{L}_0} \sqrt{\mathcal{\bar{L}}_0}}\nonumber\\
\sim&\frac{r_0^2 \sinh \left(l \sqrt{\mathcal{L}_0}\right) \sinh \left(l \sqrt{\mathcal{\bar{L}}_0}\right)}{\sqrt{\mathcal{L}_0}\sqrt{\mathcal{\bar{L}}_0}},\quad (r_0\gg1)
\end{align}
This result leads to the entanglement entropy
\begin{align}
S_{\text{EE}}
=& \frac{c}{6}\log\left(\frac{r_0^2\sinh \left(l \sqrt{\mathcal{L}_0}\right) \sinh \left(l\sqrt{\mathcal{\bar{L}}_0}\right)}{\sqrt{\mathcal{L}_0}\sqrt{\mathcal{\bar{L}}_0}}\right).
\end{align} 
If we consider the spinless black hole, i.e. $\mathcal{L}_0=\mathcal{\bar{L}}_0$, the entanglement entropy reduces to
\begin{align}
S_{\text{EE}}=&\frac{c}{3}\log\left(\frac{\beta_0}{\pi\epsilon}\sinh \left(\frac{\pi l}{\beta_0}\right) \right),\quad \beta_0=\frac{\pi}{\sqrt{\mathcal{L}_0}},
\end{align} 
where $\beta_0$ is the inverse temperature of the BTZ black hole~\cite{Aharony:1999ti,Banados:1992wn,Hubeny:2007xt}. This result coincides with the entanglement entropy of a CFT in thermal state. 
\subsection{Loops and thermal entropy}
One can also consider the Wilson loops in AdS$_3$. In this case, $W_{\mathcal{R}}(C)$ turns out to be the proper distance around the horizon, which corresponds to the black hole thermal entropy. We will then check it in the BTZ black hole. Consider the Wilson loop along the $S_1$ cycle $\theta\sim \theta+2\pi$. In contrast to the open interval case, the closed path should be smooth and hence impose the periodic boundary condition
\begin{align}
U\left(s_{f}\right) & = U(s_i), \quad P\left(s_{f}\right) = P(s_i).
\end{align}
According to \eqref{u-p}, the boundary condition for $P(s)$ implies
\begin{align}
\label{eq-p}
\left[P_{0}, R\left(s_{i}\right) R^{-1}(s_f)\right] & = 0,
\end{align}
Hence, the boundary condition for $U(s)$ implies
\begin{align}
\label{eq-u}
\exp \left(-2 \Delta \alpha P_{0}\right) & = u_{0}^{-1}\left(L^{-1}\left(s_{f}\right) L(s_i)\right) u_{0}\left(R(s_i) R^{-1}\left(s_{f}\right)\right).
\end{align}
In addition, note the relations
\begin{align}
L^{-1}\left(s_{f}\right) L(s_{i})& = \exp \left(\oint d \theta a_{\theta}\right),\\
R(s_i) R^{-1}\left(s_{f}\right) & = \exp \left(-\oint d\theta \bar{a}_{\theta}\right), \quad 
\end{align}
which are the holonomies of the connection, we can rewrite \eqref{eq-u} as
\begin{align}
\label{eq-alpha-l}
\exp \left(-2 \Delta \alpha P_{0}\right) & = u_{0}^{-1} \exp \left(2 \pi a_{\theta}\right) u_{0} \exp \left(-2 \pi \bar{a}_{\theta}\right).
\end{align}
Here we just consider the case of BTZ black hole, so that one can perform the simple integral over $\theta$.
\par
From \eqref{eq-p}, we learn that $P_0$ and $\bar{a}_{\theta}$ can be diagonalized simultaneously. If the initial value of $u_0$ is fixed, we can always choose a matrix $V$, such that $a_{\theta}$ can also be diagonalized by $u_0V$
\begin{align}
\label{eq-alpha-ll}
\exp \left(-2 \Delta \alpha \lambda_{P}\right) & = \left(u_{0} V\right)^{-1} \exp \left(2 \pi a_{\theta}\right) u_{0} V \exp \left(-2 \pi \bar{\lambda}_{\theta}\right)\nonumber\\
& =\exp \left(2 \pi {\lambda}_{\theta}\right)\exp \left(-2 \pi \bar{\lambda}_{\theta}\right),
\end{align}
where $\lambda_{P},\lambda_{\theta}$ and $\bar{\lambda}_{\theta}$ are diagonalized matrix of $P_0, a_{\theta}$ and $\bar{a}_{\theta}$. 
Contracting \eqref{eq-alpha-ll} with $L_0$, we obtain the on-shell action for the loop
\begin{align}
\label{thermal-entropy}
S_{\text{th}}&=2\pi\sqrt{2\mathcal{C}}\text{Tr}\left((\lambda_{\theta}-\bar{\lambda}_{\theta})L_0\right).
\end{align}
For the BTZ black hole, the diagonalized gauge connections are  
\begin{align}
\lambda_{\theta}=2\sqrt{\mathcal{L}_0}L_0,\quad \bar{\lambda}_{\theta}=-2\sqrt{\bar{\mathcal{L}_0}}L_0.
\end{align}
Finally, the Wilson loop gives precisely the entropy of the BTZ black hole
\begin{align}
S_{\text{th}}=2\pi\sqrt{\frac{c}{6}\mathcal{L}_0}+2\pi\sqrt{\frac{c}{6}\bar{\mathcal{L}_0}}.
\end{align}
\section{Holographic entanglement entropy in $T\bar{T}$- deformed AdS$_3$}
\label{sec:3}
We turn to investigate the entanglement entropy of $T\bar{T}$ deformed CFTs from the gravity side. In~\cite{Guica:2019nzm}, it is proposed that the holographic interpretation of $T\bar{T}$ deformed CFTs is still AdS$_3$ gravity but with the mixed boundary condition. The AdS$_3$ solutions associated with the mixed boundary condition can be obtained from the Ba\~nados geometry through a coordinate transformation. As the deformed geometry is still AdS$_3$, we prefer to work in Chern-Simons formulation. In this section, we introduce the $T\bar{T}$ deformed AdS$_3$ geometry. The holographic entanglement entropy of $T\bar{T}$ deformed CFTs can be obtained using the Wilson line technique in the deformed AdS$_3$.  
\subsection{$T\bar{T}$ deformed AdS$_3$ geometry}
We start from the general AdS$_3$ solution with a flat conformal boundary, which is called the Ba\~nados geometry~\cite{Banados:1998gg}. In Fefferman-Graham gauge, the line element reads
\begin{align}
\label{banados}
ds^2=\frac{dr^2}{r^2}+r^2\left(dzd\bar z+\frac{1}{r^2}\mathcal{L}(z)dz^2+\frac{1}{r^2}\mathcal{\bar L}(\bar z)d\bar z^2+\frac{1}{r^4}\mathcal{L}(z)\mathcal{\bar L}(\bar z)dzd\bar z\right),
\end{align}
The parameters $\mathcal{L}(z)$ and $\mathcal{\bar{L}}(\bar{z})$ are arbitrary holomorphic and antiholomorphic functions, which are related to the energy and angular momentum
\begin{align}
\label{undeformed-E}
\mathcal{L}=\frac{E+J}{2},\quad \mathcal{\bar{L}}=\frac{E-J}{2}.
\end{align}
The corresponding Chern-Simons gauge fields are  
\begin{align}
A=&\frac{dr}{r}L_0+\left(rL_{1}-\frac{1}{r}\mathcal{L}(z)L_{-1}\right)dz,\\
\bar{A}=&-\frac{dr}{r}L_0-\left(\frac{1}{r}\mathcal{\bar L}(\bar{z})L_{1}-rL_{-1}\right)d\bar z.
\end{align}
In this sense, the deformed Ba\~nados geometry can be constructed through a field-dependent coordinate transformation~\cite{Guica:2019nzm}, which reads
\begin{align}
\label{coordinate transformation}
dz=\frac{1}{1-\mu^2\mathcal{L_{\mu}\bar L_{\mu}}}(dw-\mu\mathcal{\bar L}_{\mu}d\bar w),\quad d\bar z=\frac{1}{1-\mu^2\mathcal{L_{\mu}\bar{L}_{\mu}}}(d\bar w-\mu\mathcal{L}_{\mu}dw),
\end{align}
where $\mu$ is the deformation parameter. One should note that the parameters $\mathcal{L}$ and $\mathcal{\bar{L}}$ in~\eqref{banados} would turn into $\mathcal{L}_{\mu}$ and $\mathcal{\bar{L}}_{\mu}$ under the coordinate transformation. Generally, the parameters  $\mathcal{L_{\mu}}$ and $\mathcal{\bar{L}_{\mu}}$ are different from the undeformed ones $\mathcal{L}$ and $\mathcal{\bar{L}}$. The relations between deformed parameters $\mathcal{L_{\mu}},\mathcal{\bar{L}_{\mu}}$ and undeformed parameters $\mathcal{L},\mathcal{\bar{L}}$ can be fixed by two ways. The first one is that the deformation smoothly changes the spectrum but does not change the local degeneracy of states. Therefore, in the bulk, this implies that the $T\bar{T}$ deformation does not change the horizon area of a black hole. The second one is that the deformed geometry can be transformed into the undeformed one without changing the periodicity of the spatial coordinate. Indeed, the transformation is different from the inverse of~\eqref{coordinate transformation}.
These considerations lead to
\begin{align}\label{LbarL}
\frac{\mathcal{L}_{\mu}(1-\mu \mathcal {\bar L}_{\mu})^2}{(1-\mu^2\mathcal L_{\mu}\mathcal {\bar L}_{\mu})^2}=\mathcal L,\quad
\frac{\mathcal{\bar L}_{\mu}(1-\mu \mathcal {L}_{\mu})^2}{(1-\mu^2\mathcal L_{\mu}\mathcal {\bar L}_{\mu})^2}=\mathcal{\bar L}.
\end{align}
One can turn to~\cite{Guica:2019nzm} for more details about fixing these relations.
\par
By using the coordinate transformation~\eqref{coordinate transformation}, we obtain the deformed Chern-Simons gauge fields
\begin{align}
A=&\frac{1}{r}L_0dr+\frac{1}{1-\mu^2\mathcal L_{\mu}\mathcal {\bar L}}_{\mu}\left(rL_{1}-\frac{1}{r}\mathcal{L}_{\mu}L_{-1}\right)(dw-\mu \mathcal{\bar L}_{\mu}d\bar{w}),\\
\bar{A}=&-\frac{1}{r}L_0dr-\frac{1}{1-\mu^2\mathcal L_{\mu}\mathcal {\bar L}_{\mu}}\left(\frac{1}{r}\mathcal{\bar L}_{\mu}L_{1}-rL_{-1}\right)(d\bar{w}-\mu \mathcal{L}_{\mu}dw).
\end{align}
Note that $\mathcal{L}(z)$ and $\mathcal{\bar{L}}(\bar{z})$ correspond to the charges of the solution in the Ba\~nados geometry. However, in the deformed geometry, the parameters $\mathcal{L}(z)$ and $\mathcal{\bar{L}}(\bar{z})$ do not correspond to the charges. Indeed, the deformed energy and angular momentum can be obtained from both field theory and gravity side 
\begin{align}
E_{\mu}=\frac{1}{\mu}\left(1-\sqrt{1-2\mu(\mathcal{L}+\mathcal{\bar{L}})+\mu^2(\mathcal{L}-\mathcal{\bar{L}})^2}\right),\quad J_{\mu}=J.
\end{align}
Analogous to~\eqref{undeformed-E}, we introduce the new parameters
\begin{align}
\label{q}
\mathcal{Q}=\frac{E_{\mu}+J_{\mu}}{2}=\frac{1}{2\mu}\left[1+\mu(\mathcal{L}-\bar{\mathcal{L}})-\sqrt{1-2 \mu(\mathcal{L}+\bar{\mathcal{L}})+\mu^{2}(\mathcal{L}-\bar{\mathcal{L}})^{2}}\right],\\ 
\label{qbar}
\mathcal{\bar{Q}}=\frac{E_{\mu}-J_{\mu}}{2}=\frac{1}{2\mu}\left[1-\mu(\mathcal{L}-\bar{\mathcal{L}})-\sqrt{1-2 \mu(\mathcal{L}+\bar{\mathcal{L}})+\mu^{2}(\mathcal{L}-\bar{\mathcal{L}})^{2}}\right].
\end{align}
We can regard $\mathcal{Q}$ and $\mathcal{\bar{Q}}$ as the generalized parameters of $\mathcal{L}$ and $\mathcal{\bar{L}}$ in the deformed geometry, and $\mathcal{Q}$ and $\mathcal{\bar{Q}}$ reduce to $\mathcal{L}$ and $\mathcal{\bar{L}}$ in the limit $\mu\to 0$. We find it is more convenient to parametrize the deformed gauge fields or metric in terms of these two independent charges. In terms of these charges, the Chern-Simons gauge connections are formulated as 
\begin{align}
\label{gauge field Az}
A=&\frac{dr}{r}L_0+\frac{1-\mu\mathcal{Q}}{1-\mu(\mathcal{Q}+\mathcal{\bar{Q}})}\left(r(1-\mu \mathcal{\bar{Q}})L_{1}-\frac{1}{r}\mathcal{Q}L_{-1}\right)dw\nonumber\\
&-\frac{\mu\mathcal{\bar{Q}}}{1-\mu(\mathcal{Q}+\mathcal{\bar{Q}})}\left(r(1-\mu \mathcal{\bar{Q}})L_{1}-\frac{1}{r}\mathcal{Q}L_{-1}\right)d\bar{w},\\
\label{gauge field Azbar}
\bar{A}=&-\frac{dr}{r}L_0+\frac{\mu\mathcal{Q}}{1-\mu(\mathcal{Q}+\mathcal{\bar{Q}})}\left(\frac{1}{r}\mathcal{\bar{Q}} L_{1}-r(1-\mu \mathcal{Q})L_{-1}\right)dw\nonumber\\
&-\frac{1-\mu\mathcal{\bar{Q}}}{1-\mu(\mathcal{Q}+\mathcal{\bar{Q}})}\left(\frac{1}{r}\mathcal{\bar{Q}} L_{1}-r(1-\mu \mathcal{Q})L_{-1}\right)d\bar{w},
\end{align}
\par
In the following, we prefer to use the coordinates $\theta=(w+\bar{w})/2,t=(w-\bar{w})/2$, where $t$ represents the time direction while $\theta$ denotes the spatial coordinate at the boundary with the identification $\theta\sim\theta+2\pi$. We then have
\begin{align}
\label{gauge field A}
A_{r}=\frac{1}{r}L_0,\quad A_{\theta}=&r(1-\mu \mathcal{\bar{Q}})L_{1}-\frac{1}{r}\mathcal{Q}L_{-1},\quad A_{t}=K\Big(r(1-\mu\mathcal{\bar{Q}})L_{1}-\frac{1}{r}\mathcal{Q}L_{-1}\Big),\\
\label{gauge field Abar}
\bar{A}_{r}=-\frac{1}{r}L_0,\quad \bar A_{\theta}=&\frac{1}{r}\mathcal{\bar{Q}} L_{1}-r(1-\mu \mathcal{Q})L_{-1},\quad \bar A_{t}=\bar{K}\Big(\frac{1}{r}\mathcal{\bar{Q}} L_{1}-r(1-\mu\mathcal{Q})L_{-1}\Big),
\end{align}
where
\begin{align}
K=&\frac{1+\mu(\mathcal{\bar{Q}}-\mathcal{Q})}{1-\mu(\mathcal{Q}+\mathcal{\bar Q})},\quad \bar{K}=-\frac{1-\mu(\mathcal{\bar Q}-\mathcal{Q})}{1-\mu(\mathcal{Q}+\mathcal{\bar Q})}.
\end{align}
The radial gauge~\eqref{radial-gauge} still holds for the deformed gauge fields, so that the induced gauge connections are
\begin{align}
\label{gauge field a}
a_{\theta}=&(1-\mu\mathcal{\bar{Q}})L_{1}-\mathcal{Q}L_{-1},\quad a_{t}=K\Big((1-\mu\mathcal{\bar{Q}})L_{1}-\mathcal{Q}L_{-1}\Big),\\
\label{gauge field abar}
\bar a_{\theta}=&\mathcal{\bar{Q}} L_{1}-(1-\mu \mathcal{Q})L_{-1},\quad \bar a_{t}=\bar{K}\Big(\mathcal{\bar{Q}} L_{1}-(1-\mu \mathcal{Q})L_{-1}\Big).
\end{align}
In addition, we can also write down the deformed 
\begin{align}
\label{deformed ads}
ds^2=&\frac{dr^2}{r^2}+\frac{1}{r^2(1-\mu(\mathcal{Q}+\mathcal{\bar{Q}}))^2}\times\nonumber\\
&\Big(\mathcal{Q}(1-\mu \mathcal{Q})(1-\mu r^2)dw+\Big(\mu\mathcal{Q}\mathcal{\bar{Q}}+r^2(1-\mu\mathcal{Q})(1-\mu\mathcal{\bar{Q}})\Big)d\bar{w}\Big)\times\nonumber\\
&\Big(\mathcal{\bar{Q}}(1-\mu\mathcal{\bar{Q}})(1-\mu r^2)d\bar{w}+\Big(\mu \mathcal{Q}\mathcal{\bar{Q}}+r^2(1-\mu\mathcal{Q})(1-\mu\mathcal{\bar{Q}})\Big)dw\Big).
\end{align}
We will use the deformed geometry to calculate the holographic entanglement entropy in the $T\bar{T}$ deformed CFTs. For simplicity, we just consider the constant charges $\mathcal{Q}$ and $\mathcal{\bar{Q}}$, namely we work in $T\bar{T}$ deformed BTZ black hole. 
\subsection{$T\bar{T}$-deformed holographic entanglement entropy}
For the $T\bar{T}$-deformed AdS$_3$, the metric still satisfies the Einstein equation or flat connection condition in the Chern-Simons theory although it takes a complicated form. In the Poincar\'e AdS$_3$, the Wilson line would produce a back-reaction in the bulk geometry. The back-reaction would then lead to a conical defect on the ending points of Wilson line, which generates the $n$-sheet manifold on the boundary. According to the replica trick on the boundary field theory, the Wilson line exactly leads to the entanglement entropy. One can turn to Appendix~\ref{app:defect} for details. We can always transform the $T\bar{T}$-deformed AdS$_3$ solution into the Poincar\'e form~\cite{Rooman:2000ei,Krasnov:2001cu}. However, the temperature (the period of Euclidean time) in deformed AdS$_3$ is different from the original one. The crucial point is that we have to identify the deformed temperature and length of interval on the boundary under $T\bar{T}$ deformation. We will treat these considerations in more details and obtain the $T\bar{T}$ deformed holographic entanglement entropy in this section. 
\par
Now, we can use the Wilson line technique to calculate the holographic entanglement entropy in $T\bar{T}$-deformed AdS$_3$. First of all, we can give a glance at the Poincar\'e AdS$_3$, which turns out correspond to the zero temperature entanglement entropy. In Fefferman-Graham gauge, the Poincar\'e AdS$_3$ can be written as Ba\~nados geometry~\eqref{banados} with $\mathcal{L}$ and $\bar{\mathcal{L}}$ vanish. In this case, the bulk geometry is the same as the undeformed one, so the zero temperature entanglement entropy remains unchanged. This result coincides with the perturbative calculation in field theory and cutoff perspective in the bulk~\cite{Chen:2018eqk,Jeong:2019ylz}.
\par
We then consider the deformed BTZ black hole, in which the charges $\mathcal{Q}$ and $\mathcal{\bar{Q}}$ are constants. For the deformed geometry, on a time slice, we obtain 
\begin{align}
L\left(r,\theta,t=0\right) & = \exp \left(-\ln r L_{0}\right) \exp \left(-\int_{x_{0}}^{x} d x^{i} a_{i}\right)\nonumber\\
&= \exp \left(-\ln r L_{0}\right)\exp \left(-(1-\mu\mathcal{\bar{Q}})\theta L_{1}+\mathcal{Q}\theta L_{-1}\right),\\
R\left(r,\theta,t=0\right) & = \exp \left(\int_{x_{0}}^{x} d x^{i} \bar{a}_{i}\right) \exp \left(-\ln r L_{0}\right) \nonumber\\
 & =\exp \left(\mathcal{\bar{Q}}\theta L_{1}-(1-\mu \mathcal{Q})\theta L_{-1}\right) \exp \left(-\ln r L_{0}\right). 
\end{align}
As the deformed geometries are still  AdS$_3$ solution, we use the boundary condition for $U(s)$
\begin{align}
\label{u-tt}
U(s_i)=\textbf{1},\quad U(s_f)=\textbf{1},
\end{align}
as well as the same boundary conditions for the ending points of the Wilson line
\begin{align}
&r(s_i)=r(s_f)=r_0,\\
&\Delta\theta=\theta(s_f)-\theta(s_i)=l.
\end{align} 
We should point out that the boundary condition for $U$ is actually the unique choice because of the Lorentz invariance at the boundary~\cite{Ammon:2013hba,Castro:2015csg}. As the $T\bar{T}$ deformation does not break Lorentz invariance, we can use the same boundary condition~\eqref{u-tt} for $U$.
It seems that $l$ is just the length of the interval in the deformed boundary. But it equals to the deformed length of interval, because the length is defined in the $(w,\bar{w})$ coordinates. 
\par
Using the gauge transformation \eqref{u-p}, one can get the solution $U(s)$ for the Wilson line coupled to the deformed gauge fields. The boundary condition for $U(s)$ and ending points boundary condition for the Wilson line imply
\begin{align}
&\text{Tr}\Big(\left(R(s_i)L(s_i)\right)\left(R\left(s_{f}\right) L\left(s_{f}\right)\right)^{-1}\Big)\nonumber\\
=&2 \cosh \left(l \sqrt{\mathcal{\bar{Q}}\left(1-\mu\mathcal{Q}\right)}\right) \cosh \left(l\sqrt{\mathcal{Q}(1-\mu \mathcal{\bar{Q}})}\right)\nonumber\\
&+\frac{r_0^2 \sqrt{\mathcal{\bar{Q}} (1-\mu\mathcal{Q})} \sqrt{\mathcal{Q}(1-\mu\mathcal{\bar{Q}})} \sinh \left(l\sqrt{\mathcal{\bar{Q}} (1-\mu\mathcal{Q})}\right) \sinh \left(l \sqrt{\mathcal{Q}(1-\mu\mathcal{\bar{Q}})}\right)}{\mathcal{Q}\mathcal{\bar{Q}}}\nonumber\\
&+\frac{\mathcal{Q}\mathcal{\bar{Q}} \sinh \left(l\sqrt{\mathcal{\bar{Q}}(1-\mu\mathcal{Q})}\right) \sinh \left(l\sqrt{\mathcal{Q}(1-\mu \mathcal{\bar{Q}})}\right)}{r_0^2 \sqrt{\mathcal{\bar{Q}} (1-\mu\mathcal{Q})} \sqrt{\mathcal{Q}(1-\mu\mathcal{\bar{Q}})}}\nonumber\\
\sim&\frac{r_0^2 \sqrt{\mathcal{\bar{Q}} (1-\mu  \mathcal{Q})} \sqrt{\mathcal{Q}(1-\mu\mathcal{\bar{Q}})} \sinh \left(l\sqrt{\mathcal{\bar{Q}} (1-\mu\mathcal{Q})}\right) \sinh \left(l \sqrt{\mathcal{Q}(1-\mu \mathcal{\bar{Q}})}\right)}{\mathcal{Q}\mathcal{\bar{Q}}}.
\end{align}
In the last step, we consider the $r_0\gg 1$ limit. It is straightforward to get the holographic entanglement entropy for $T\bar{T}$ deformation
\begin{align}\label{HEEwil}
S_{\text{EE}}=&\sqrt{2\mathcal{C}}\cosh^{-1}\left(\frac{r_0^2 \sqrt{\mathcal{\bar{Q}} (1-\mu  \mathcal{Q})} \sqrt{\mathcal{Q}(1-\mu\mathcal{\bar{Q}})} \sinh \left(l\sqrt{\mathcal{\bar{Q}} (1-\mu\mathcal{Q})}\right) \sinh \left(l \sqrt{\mathcal{Q}(1-\mu \mathcal{\bar{Q}})}\right)}{2\mathcal{Q}\mathcal{\bar{Q}}}\right)\nonumber\\
\sim&\frac{c}{6}\log\left(\frac{r_0^2 \sqrt{\mathcal{\bar{Q}} (1-\mu  \mathcal{Q})} \sqrt{\mathcal{Q}(1-\mu\mathcal{\bar{Q}})} \sinh \left(l\sqrt{\mathcal{\bar{Q}} (1-\mu\mathcal{Q})}\right) \sinh \left(l \sqrt{\mathcal{Q}(1-\mu \mathcal{\bar{Q}})}\right)}{\mathcal{Q}\mathcal{\bar{Q}}}\right).
\end{align} 
\par
If the original geometry is non-rotating BTZ black hole, namely $\mathcal{Q}=\mathcal{\bar{Q}}$, the deformed entanglement entropy becomes
\begin{align}
S_{\text{EE}}
=&\frac{c}{3}\log \left(\frac{r_0\sqrt{\mathcal{Q}(1-\mu\mathcal{Q})}\sinh\left(l\sqrt{\mathcal{Q}(1-\mu\mathcal{Q})}\right)}{\mathcal{Q}}\right).
\end{align}
We then want to identify the deformed temperature. For the deformed BTZ black hole, the temperature can be obtained by analyzing the period of Euclidean time, which is discussed in the next section (equation \eqref{temdef}). We quote the result here
\begin{align}
\label{deform-tem-1}
\beta=\frac{1}{T}=\frac{\pi(1-2\mu\mathcal{Q})}{\sqrt{\mathcal{Q}(1-\mu\mathcal{Q})}}.
\end{align}
This temperature can also be derived using the first law of thermodynamics, and we will show it in section~\ref{TT-loop}. For the limit $\mu\to 0$, the temperature reduces to the BTZ black hole temperature. The length of interval $l$ is already the deformed one, which can be seen from the coordinate transformation~\eqref{coordinate transformation} on a time slice. In terms of the deformed temperature, we can express the entanglement entropy as
\begin{align}
\label{HEEttbar}
S_{\text{EE}}=\frac{c}{3}\log \left(\frac{\sqrt{\beta^2+4\mu\pi^2}+\beta}{2\pi\epsilon}\sinh \left(\frac{\pi  l}{ \sqrt{\beta ^2+4\mu \pi^2 }}\right)\right).
\end{align}
This is actually the $T\bar{T}$ deformed entanglement entropy obtained from the holographic approach. For $\mu =0$, the deformed entanglement entropy reduces to the familiar entanglement entropy of CFT at finite temperature. For the small $\mu$, we can obtain the perturbative result
\begin{align}
S_{\text{EE}}=\frac{c}{3}\log \left(\frac{\beta}{\pi \epsilon}\sinh \left(\frac{\pi l}{\beta}\right)\right)+\frac{\mu c}{3}\left(\frac{\pi^2}{\beta^2}-\frac{2\pi^3 l}{\beta^3}\coth \left(\frac{\pi l}{\beta}\right)\right)+O(\mu^2).
\end{align}
In the ``low temperature'' limit $\beta\gg l$, up to the first order, the entanglement entropy becomes
\begin{align}
S_{\text{EE-low}}
=&\frac{c}{3}\log \left(\frac{\beta}{\pi\epsilon}\sinh\left(\frac{\pi l}{\beta}\right)\right)+\frac{\mu c}{3}\left(\frac{\pi^2}{\beta^2}\right)+O(\mu^2).
\end{align}
In the ``high temperature'' limit $\beta\ll l$, the first order corrected entanglement entropy is
\begin{align}
\label{TT-EE-high}
S_{\text{EE-high}}
=&\frac{c}{3}\log \left(\frac{\beta}{\pi\epsilon}\sinh\left(\frac{\pi l}{\beta}\right)\right)-\frac{2\mu c}{3}\frac{\pi^3 l}{\beta^3}\coth \left(\frac{\pi l}{\beta}\right)+O(\mu^2).
\end{align}
The ``high temperature'' result coincides with the result obtained from both boundary field side and AdS$_3$ with cutoff perspective~\cite{Chen:2018eqk,Jeong:2019ylz}\footnote{Note that our convention is different from Ref. \cite{Chen:2018eqk}. In \cite{Chen:2018eqk}, the deformation parameter is related to the radial cutoff $r_c^2=\frac{6}{\mu\pi c}$, while we have $r_c^2=\frac{1}{\mu}$ in this paper. Therefore, if one replaces $\mu$ by $\frac{\mu\pi c}{6}$, the equation~\eqref{TT-EE-high} becomes
\begin{align}
S_{\text{EE-high}}
=&\frac{c}{3}\log \left(\frac{\beta}{\pi\epsilon}\sinh\left(\frac{\pi l}{\beta}\right)\right)-\frac{\mu\pi^4 c^2 l}{9\beta^3}\coth \left(\frac{\pi l}{\beta}\right).
\end{align}
which is exactly the result in~\cite{Chen:2018eqk}.}. We apply the Wlison line approach to the $T\bar{T}$-deformed AdS$_3$ and obtain the holographic entanglement entropy formula, which agree with the perturbation results. However, the ``low temperature'' result is different from the cutoff AdS$_3$ perspective. 
\par
We are more interested in the non-perturbative result. In order to make sure the entanglement entropy is real, we have
\begin{equation}
-\frac{\beta^2}{4\pi^2}<\mu,
\end{equation} 
which means the holographic description maybe loses when $\mu$ out of this region. For $\mu>0$ the entanglement entropy is always real. In the following discussion, we just consider the $\mu>0$ case, which also corresponds to the cutoff perspective. For a fixed temperature, we can consider the entanglement entropy for large deformation parameter 
\begin{align}
\label{large-deformation}
S_{\text{EE}}=\frac{c}{3}\log \left(\frac{l}{2 \epsilon }\right)+\frac{\beta  c }{6 \pi}\frac{1}{\sqrt{\mu}}+\left(\frac{c l^2}{72}-\frac{\beta ^2 c}{24 \pi ^2}\right)\frac{1}{\mu}+O\left(\frac{1}{\mu}\right).
\end{align}
The leading order coincides with the entanglement entropy of the zero temperature CFT with the length of interval $l/2$. Therefore the holographic entanglement entropy has nothing to do with the temperature for large deformation parameter. However, we can not directly see this feature from the cutoff perspective, the reason is that the cutoff surface would move into the horizon for large deformation parameter at a fixed finite temperature. In order to see this feature from cutoff perspective, we have to consider the holographic entanglement entropy at the high temperature limit, so that we can keep the cutoff surface outside the horizon even for large deformation parameter. In this case, we first take the high temperature limit and then consider the large deformation parameter limit. For the high temperature limit, the entanglement entropy~\eqref{deform-tem-1} becomes
\begin{align}
S_{\text{EE}}=\frac{c}{3}\log \left(\frac{\sqrt{\mu}}{\epsilon} \sinh \left(\frac{l}{2 \sqrt{\mu}}\right)\right)+\frac{c\beta}{6 \pi  \sqrt{\mu}}+O(\beta^2) .   
\end{align}
This formula leads to~\eqref{large-deformation} for the large deformation parameter as well. These results imply the $T\bar{T}$ deformation behaves like the free theory at the large $\mu$ limit. The similar feature was also found in~\cite{Chakrabarti:2020pxr,Chakrabarti:2020dhv}, in which the authors shown that at the level of the equations of motion the left- and right-chiral sectors of $T\bar{T}$ deformed free theories are decoupled when the deformation parameter is sent to infinity. 
\par
For the holographic entanglement entropy in BTZ black hole, it turns out that there will be a phase transition when the interval becomes larger~\cite{Ryu:2006bv,Ryu:2006ef}. The similar phase transition are also exist in $T\bar{T}$-deformed BTZ black hole. As the size of interval becomes larger, we find the deformed holographic entanglement entropy becomes
\begin{align}
S_{\text{EE}}\sim\frac{c \pi l}{3}\frac{1}{\sqrt{\beta^2+4\pi^2\mu}},
\end{align}
which is proportional to the thermal entropy~\eqref{thermal-entropy-tt-b}. This result can be interpreted as follows. In the presence of horizon, the RT surface (geodesic line) would approach the horizon and cover a part of the horizon when the size of interval becomes larger. Since the geodesic lines could wrap different parts of the horizon, there are two geodesics connecting the endpoints of the interval.  The one being homologous to the boundary interval should be chosen to evaluate the holographic entanglement entropy, and the other geodesic is used to evaluate the entanglement entropy for the complementary interval
\begin{align}
S'_{\text{EE}}=\frac{c}{3}\log \left(\frac{\sqrt{\beta^2+4\mu\pi^2}+\beta}{2\pi\epsilon}\sinh \left(\frac{\pi(2\pi-l)}{ \sqrt{\beta ^2+4\mu \pi^2 }}\right)\right).    
\end{align}
However, if $l>\pi$ the $S'_{\text{EE}}$ is smaller than $S_{\text{EE}}$,  but $S'_{\text{EE}}$ is not homologous to the interval $l$. Similar to the BTZ black hole~\cite{Blanco:2013joa}, we can construct another disconnected extremal surface by combining $S'_{\text{EE}}$ and thermal entropy~\eqref{thermal-entropy-tt-b}. Finally, the deformed entanglement entropy for a general interval is given by
\begin{align}
S=\text{min}\{S_{\text{EE}},S'_{\text{EE}}+S_{\text{th}}\}.    
\end{align}
Clearly, there is a phase transition between these two settings of extremal surface. In the high temperature limit and small deformation parameter, it turns out that the phase transition occurs at
\begin{align}
l\sim 2\pi-\frac{\sqrt{\beta ^2+4\mu \pi^2 }}{2\pi}\log 2.
\end{align}
We show the phase transition of the holographic entanglement entropy is preserved under $T\bar{T}$ deformation.
\par
Moreover, the Casini-Huerta entropic $c$-function~\cite{Casini:2006es} for the $T\bar{T}$ deformed entanglement entropy is
\begin{align}
C(l,\mu)=l\frac{dS_{\text{EE}}}{dl}=\frac{\pi  c l }{3 \sqrt{\beta ^2+4 \pi ^2 \mu }}\coth \left(\frac{\pi  l}{\sqrt{\beta ^2+4 \pi ^2 \mu }}\right),
\end{align}
which is always positive, and does not depend on the ultraviolet regulator. We also find that 
\begin{align}
\frac{\partial C(l,\mu)}{\partial l}=\frac{ \pi  c}{3} \left(\frac{\coth \left(\frac{\pi  l}{\sqrt{\beta ^2+4 \pi ^2 \mu }}\right)}{\sqrt{\beta ^2+4 \pi ^2 \mu }}-\frac{\pi  l \text{csch}^2\left(\frac{\pi  l}{\sqrt{\beta ^2+4 \pi ^2 \mu }}\right)}{\beta ^2+4 \pi ^2 \mu }\right)\geq 0,    
\end{align}
which implies the entropic $c$-function is non–decreasing along the renormalization group flow towards the ultraviolet. The similar result was also found in single trace $T\bar{T}$ deformation~\cite{Asrat:2019end}.
\subsection{Thermal entropy}
\label{TT-loop}
The thermal entropy of the deformed BTZ black hole can also be calculated from the Wilson loop. As discussed in section 2.4, the thermal entropy can be obtained by diagonalizing the induced gauge connections $a_{\theta}$ and $\bar{a}_{\theta}$ in \eqref{gauge field a} and \eqref{gauge field abar}. 
For the deformed BTZ black hole, the diagonalized gauge connections read
\begin{align}
\lambda_{\theta}&=2\sqrt{\mathcal{Q}(1-\mu\mathcal{\bar{Q}})}L_0=2\sqrt{\mathcal{L}}L_0,\\
\quad \bar{\lambda}_{\theta}&=-2\sqrt{\mathcal{\bar{Q}}(1-\mu\mathcal{Q})}L_0=-2\sqrt{\bar{\mathcal{L}}}L_0.
\end{align}
Finally, according to~\eqref{thermal-entropy}, we obtain the thermal entropy 
\begin{align}
S_{\text{th}}&=2\pi\sqrt{\frac{c}{6}\mathcal{L}}+2\pi\sqrt{\frac{c}{6}\bar{\mathcal{L}}},
\end{align}
which is the same as the BTZ black hole entropy. This result means the black hole entropy does not change under the $T\bar{T}$ deformation. On the field theory side, the degeneracy of states does not change under the $T\bar{T}$ flow.
\par
For the deformed theory, the thermal entropy should be expressed in terms of the deformed energy. In case of $\mathcal{Q}=\mathcal{\bar{Q}}$, the entropy can be written as 
\begin{align}
S_{\text{th}}&=4\pi\sqrt{\frac{c}{6}\mathcal{Q}(1-\mu\mathcal{Q})}=2\pi\sqrt{\frac{c}{6}E_{\mu}(2-\mu E_{\mu})},
\end{align} 
which agrees with the result in~\cite{McGough:2016lol}. The thermal entropy can help us to define the temperature in the $T\bar{T}$-deformed theory. In fact, according to the first law of thermodynamics, the temperature can be determined by 
\begin{align}
\label{deform-tem-2}
T=\frac{\partial E_{\mu}}{\partial S_{\text{th}}}=\sqrt{\frac{6}{c}}\frac{ \sqrt{\mathcal{Q} (1-\mu \mathcal{Q})}}{\pi(1-2 \mu\mathcal{Q})}\sim \frac{ \sqrt{\mathcal{Q} (1-\mu \mathcal{Q})}}{\pi(1-2 \mu\mathcal{Q})},
\end{align}
where we have used the convention $k={c}/{6}=1$ in the definition of temperature. This is actually the temperature we have used in~\eqref{deform-tem-1}. The black hole thermal entropy can also be written in terms of temperature
\begin{align}
\label{thermal-entropy-tt-b}
S_{\text{th}}=\frac{c}{3}\frac{2\pi^2}{\sqrt{\beta^2+4\pi^2\mu}}.
\end{align}
\subsection{Two intervals entanglement entropy}
\begin{figure}[htb] 
 \center{\includegraphics[width=10cm]  {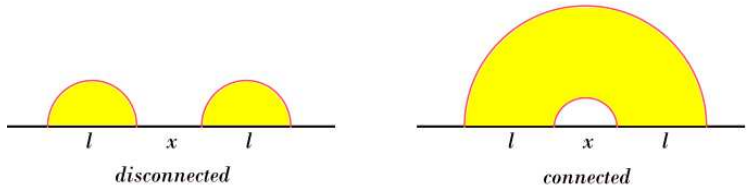}} 
 \caption{\label{p1} The two minimal surfaces for the two intervals boundary region. We consider the two intervals have the same length $l$ separated by $x$. The left is the disconnected case, and the right is the connected case.} 
\end{figure}

We proceed to consider the entanglement entropy of the system consists of two disjoint intervals. For the single interval case, we have shown that the entanglement entropy is the Wilson line or length of geodesic in AdS$_3$ with ending points on the spatial infinity boundary for both Brown-Henneaux boundary condition and mixed boundary condition. According to Ryu-Takayanagi's proposal~\cite{Ryu:2006bv,Ryu:2006ef}, we have two choices for how to draw the geodesics that end on the ending points of two intervals, which are shown in Figure~\ref{p1}. For each choice, the two intervals entanglement entropy decouples into a sum of single interval cases. 
The two intervals holographic entanglement entropy should be the minimal one of them
\begin{align}
S_{\text{EE-2}
}=\text{min}\{S_{\text{dis}},S_{\text{con}}\}.
\end{align}
This implies that there are two phases of the entanglement entropy. It turns out that there actually exist a phase transition between the connected and disconnected phase~\cite{Hartman:2013mia}. 
\par
We first brief review the zero temperature entanglement entropy of two disjoint intervals. We assume the two intervals have the same length $l$ separated by $x$, described in Figure~\ref{p1}. Then the difference between two phases is
\begin{align}
\Delta S=S_{\text{dis}}-S_{\text{con}}=\frac{c}{3}\log\left(\frac{l^2}{x(2l+x)}\right).
\end{align}
One can find the phase transition critical point is determined by the cross-ratio
\begin{align}
\eta=\frac{l^2}{(l+x)^2}=\frac{1}{2}\quad \text{or}\quad \frac{x}{l}=\sqrt{2}-1.
\end{align} 
For the finite temperature case, the similar phase transition was shown in ~\cite{Headrick:2010zt,Fischler:2012uv}. However, there is no quantity like cross-ratio to illustrate the critical point.
\par
Now we would like to investigate the similar feature for the $T\bar{T}$ deformed entanglement entropy. For the different choices of Wilson lines or RT surfaces, we have
\begin{align}
S_{\text{dis}}
=&\frac{c}{3}\log \left(\frac{\pi ^2 \mu +\frac{1}{2} \beta  \left(\sqrt{\beta ^2+4 \pi ^2 \mu }+\beta \right)}{\pi^2  \epsilon^2 }\sinh^2 \left(\frac{\pi  l}{ \sqrt{\beta ^2+4\mu \pi^2 }}\right)\right),\\
S_{\text{con}}
=&\frac{c}{3}\log \left(\frac{\pi ^2 \mu +\frac{1}{2} \beta  \left(\sqrt{\beta ^2+4 \pi ^2 \mu }+\beta \right)}{\pi^2  \epsilon^2 }\sinh \left(\frac{\pi x}{ \sqrt{\beta ^2+4\mu \pi^2 }}\right)\sinh\left(\frac{\pi(2l+x)}{ \sqrt{\beta ^2+4\mu \pi^2 }}\right)\right).
\end{align}
The two intervals entanglement entropy is the minimal one of them. In order to determine which is the minimal one and under what conditions the phase transition happens, we consider the difference between two RT surfaces
\begin{align}
\label{delta-S-0}
\Delta S=&S_{\text{dis}}-S_{\text{con}}
=\frac{c}{3}\log\left(\frac{\sinh^2\left(\frac{\pi l}{ \sqrt{\beta ^2+4\mu \pi^2 }}\right)}{\sinh\left(\frac{\pi x}{ \sqrt{\beta ^2+4\mu \pi^2 }}\right)\sinh\left(\frac{\pi(2l+x)}{ \sqrt{\beta ^2+4\mu \pi^2 }}\right)}\right).
\end{align} 
This quantity is also related to the mutual information between two disjoint subsystems. From~\eqref{delta-S-0}, we learn that $\Delta S$ behaves like the undeformed one but with different temperature. We first consider the low temperature and high temperature limit. For the low temperature limit $\beta\gg 1$, we have
\begin{align}
\Delta S=\frac{c}{3}\log \left(\frac{l^2}{x(2l+x)}\right)+O\left(1/\beta^2\right).
\end{align}
The leading order is exactly the zero temperature case. The phase transition occurs at $x/l=\sqrt{2}-1$ and does not depend on the deformation parameter.
For the high temperature limit $\beta\ll 1$, we have
\begin{align}
\Delta S=\frac{c}{3}\log \left(\frac{\cosh \left(\frac{l}{\sqrt{\mu }}\right)-1}{\cosh \left(\frac{l+x}{\sqrt{\mu }}\right)-\cosh \left(\frac{l}{\sqrt{\mu }}\right)}\right)+O\left(\beta^2\right).
\end{align}
In this case, the critical point depends on the deformation parameter.
\par
\begin{figure}[htb] 
 \center{\includegraphics[width=10cm]{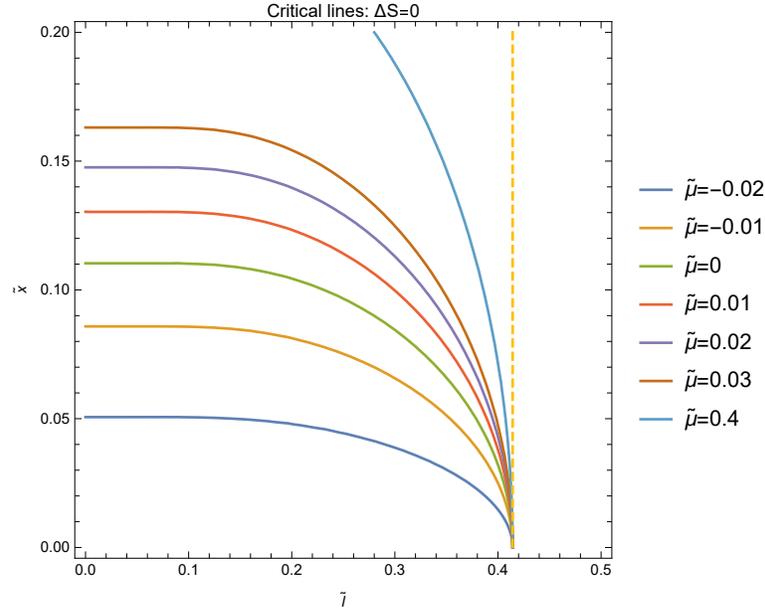}} 
 \caption{\label{p2} Plot the critical lines $\Delta S=0$ in $\tilde{l}-\tilde{x}$ plane for different deformation parameters. The critical lines separate the connected phase (left side) and disconnected phase (right side). The green line corresponds to the undeformed case. The dashed line denotes the zero temperature  critical line $\tilde{l}=\sqrt{2}-1$. The critical lines tend to the zero temperature case with the increase of deformation parameter.} 
\end{figure}

 
We find it is convenient to introduce the following parameters
\begin{align}
\tilde{l}=\frac{x}{l},\quad \tilde{x}=\frac{x}{\beta},\quad \tilde{\mu}=\frac{\mu}{\beta^2}.
\end{align}
In terms of the new parameters, the $\Delta S$ reduces to
\begin{align}
\Delta S=\frac{c}{3}\log\left(\frac{\sinh^2\left(\frac{\pi\tilde{x}}{\tilde{l} \sqrt{1+4\tilde{\mu}\pi^2 }}\right)}{\sinh\left(\frac{\pi\tilde{x}}{\sqrt{1+4\tilde{\mu} \pi^2 }}\right)\sinh\left(\frac{\pi(2+\tilde{l})\tilde{x}}{\tilde{l}\sqrt{1+4\tilde{\mu}\pi^2 }}\right)}\right),
\end{align}
in which the temperature is implicit. We plot the critical lines $\Delta S=0$ in $(\tilde{l},\tilde{x})$ plane for different deformation parameters in Figure~\ref{p2}.
Then we consider some special limit about the critical lines. For $\tilde{x}\ll 1$, we have
\begin{align}
\Delta S=\frac{c}{3}\left[\log \left(\frac{1}{\tilde{l}^2+2 \tilde{l}}\right)-\frac{ \pi ^2(\tilde{l}+1)^2\tilde{x}^2}{3\tilde{l}^2 \left(1+4\tilde{\mu}\pi ^2\right)}\right]+O\left(\tilde{x}^3\right).
\end{align}
The leading order is just the zero temperature case and also does not depend on the deformation parameter. This result can be seen from Figure~\ref{p2} that the critical lines coincide with the zero temperature one for small $\tilde{x}$.
\par 
It is interesting to investigate the $\mu$ dependence of phase transition. For the small $\tilde{\mu}$, there is actually exists a phase transition, which has been discussed in~\cite{Jeong:2019ylz} using the perturbative method. We can also see from Figure~\ref{p2} the critical line is around the undeformed case for both $\tilde{\mu} <0$ and $\tilde{\mu}>0$. For the $\tilde{\mu}\gg1$ region, we have
\begin{align}
\Delta S=\frac{c}{3}\log \left(\frac{1}{\tilde{l}^2+2\tilde{l}}\right)-\frac{c (\tilde{l}+1)^2 \tilde{x}^2}{36\tilde{l}^2 \tilde{\mu} }+O(1/\tilde{\mu}^2).
\end{align}
The leading order is just the zero temperature case. One can also see from Figure~\ref{p2} that the critical lines would become the zero temperature one as the increase of deformation parameters. This result implies the $T\bar{T}$ deformed theory becomes a decoupled free theory for large $\mu$ limit~~\cite{Chakrabarti:2020pxr,Chakrabarti:2020dhv}. 
\par
These results show that there still exist the phase transition for two intervals entanglement entropy under $T\bar{T}$ deformation. The transition point is depends on the deformation parameter. The $T\bar{T}$ deformation does not introduce new phases. For large deformation parameter, the critical point is the same as zero temperature CFT case, it would be interesting to study this feature from the field theoretic results. 
\section{Geodesic line method}
\label{sec:4}
In this section we re-compute the holographic entanglement entropy in BTZ background with mix boundary condition using RT formula, i.e., identifying the holographic entanglement entropy as the geodesic distance. The results turn out to be consistent with the computation via Wilson line method. 

The metric of BTZ black hole with mass $M$ and angular momentum $J$ takes the form (\ref{BTZme}). 
\footnote{We follow the convention in \cite{Guica:2019nzm}, and set $4\pi G=1,l=1$ and $R=2\pi$ (periodicity of spatial dimension) in their paper. We also use $r$ which is related with the radial coordinate $\rho$ in \cite{Guica:2019nzm} as $r^2=1/\rho$. The cutoff in \cite{Guica:2019nzm} locates at $\rho=\rho_c=\m$, then in $r$-coordinate, $r_0=r_c=1/\sqrt{\m}$.}
For simplicity we consider the case where the black hole being static $J=0$. It follows from (\ref{LbarL}) that the deformed parameters $\L_\m, \bar{\L}_\m$ are constant and  satisfy
\be 
\L_\m=\bar{\L}_\m=\frac{1-\m M\pm \sqrt{1-2\m M}}{M \m^2},
\ee
where only the solution with ``-'' is well defined in $\m\to 0$ limit. We start from the following metric
\be\ba \label{me}
ds^2=&\frac{dr^2}{r^2}+r^2\Big(dzd\bar{z}+\frac{1}{r^2}(\L_\m dz^2+\bar{\mathcal{L}}_\m d\bar{z}^2)+\frac{1}{r^4}\L_\m\bar{\L}_\m dzd\bar{z}\Big),
\ea\ee
in which we have replaced the $\L,\bar\L$ by $\L_\m, \bar{\L}_\m$ in the BTZ black hole solution, so that we can obtain the deformed BTZ only by using the coordinate transformation. Let $z=x+iy$, and define 
\be
r=\sqrt{\L_\m} e^\rho,\quad x=  \frac{\bar{x}}{\sqrt{4\L_\m}},\quad y= \frac{\bar{y}}{\sqrt{4\L_\m}},
\ee
then the metric becomes the global AdS$_3$
\be\ba \label{me}
ds^2
=&d\rho^2+\cosh^2\rho d\bar{x}^2+\sinh^2\rho d\bar{y}^2,
\ea\ee
where $\bar{y}$  is treated as the Euclidean time and  $\bar{x}$ the spatial coordinate. The requirement of no conical singularity in $\rho-\bar{y}$ plane implies the identification  $\bar{y}\sim \bar{y}+2\pi$, where the periodicity is related with the temperature for BTZ black hole. It is convenient to work in embedding coordinate 
\be\ba 
Y^0=&\cosh\rho \cosh \bar{x},~~Y^3=\cosh\rho \sinh \bar{x},\\
Y^1=&\sinh\rho\sin \bar{y},~~Y^2=\sinh\rho \cos \bar{y}.
\ea\ee
In this coordinate system the BTZ black hole is a hypersurface $-(Y^0)^2+(Y^3)^2+(Y^1)^2+(Y^2)^2=-1$ in the background $ds^2=-d(Y^0)^2+d(Y^1)^2+d(Y^2)^2+d(Y^3)^2$.  The geodesic distant $d$ between two points $Y_1^a,Y_2^b$ is simply computed by
\be \label{geod}
\cosh d=-Y_1\cdot Y_2=Y^0_1Y^0_2-Y^1_1Y^1_2-Y^2_1Y^2_2-Y^3_1Y^3_2.
\ee

The deformed metric corresponding to $T\bar{T}$ deformation can be obtained by transformation of
\be \label{tra}
dz =\frac{1}{1-\m^2 \L_\m\bar{\L}_\m}(dw-\m \bar{\L}_\m d\bar{w}),~~d\bar{z}=\frac{1}{1-\m^2 \mathcal{L}_\m\bar{\L}_\m}(d\bar{w}-\m \L_\m dw).
\ee
In the present case, (\ref{tra}) can be solved straightforwardly as  
\be 
 z =\frac{1}{1-\m^2 \L_\m\bar{\L}_\m}( w-\m \bar{\L}_\m  \bar{w}),~~ \bar{z}=\frac{1}{1-\m^2 \mathcal{L}_\m\bar{\L}_\m}( \bar{w}-\m \L_\m  w).
\ee
And its inverse 
\be \label{wzt}
w=z+\m \bar{\L}_\m \bar{z},~~\bar{w}=\m\L_\m z+\bar{z},
\ee
where $w=\theta+ it,\bar{w}=\theta-it$. From the periodicity of $\bar{y}$ discussed above, we can work out the period of $t$, which is 
\be\ba \label{temdef}
t\sim  t+\frac{2\pi (1-\m \L _\m)}{\sqrt{4\L_\m}}=t+\beta,~~\beta= 
 \frac{\pi(1-2\m \Q)}{\sqrt{\Q(1-\m \Q)}},
\ea\ee
where the $\beta$ is the inverse temperature of deformed black hole, as well as the inverse temperature of the $T\bar{T}$ deformed CFT.

To compute the HEE of a single interval, we consider two ending points  on the boundary locate at $(r_1,t_1,\theta_1)=(\sqrt{\L_\m}e^{\rho_0},0,0)$ and  $(r_2,t_2,\theta_2)=(\sqrt{\L_\m}e^{\rho_0},0,l)$ respectively. Then $w_1=\bar{w}_1=0$, $w_2=\bar{w}_2=l$
\be 
z_1=\bar{z}_1=0,~~
z_2 =\bar{z}_2=\frac{l}{1+\m\L_\m}.
 \ee
In terms of embedding coordinates 
\be\ba 
Y_1^0=&\cosh\rho_0 ,~~Y_1^3=0,~~Y^1_1=&0,~~Y_1^2=\sinh\rho_0, 
\ea\ee
and 
\be\ba 
Y^0_2=&\cosh\rho_0 \cosh \sqrt{4\L_\m}z_2,~~Y^3_2=\cosh\rho \sinh \sqrt{4\L_\m}z_2,~~Y^1_2=0,~~Y^2_2=\sinh\rho_0.
\ea\ee
Finally using (\ref{geod}), the geodesic distance between the points is 
\be\ba 
\cosh d=&\cosh^2\rho_0  \cosh \sqrt{4\L_\m}z_2-\sinh^2\rho_0\\
=& \frac{\Q}{2r_0^2(1-\m \Q)}\sinh^2l\sqrt{\Q(1-\m \Q)}+\cosh^2l\sqrt{\Q(1-\m \Q)}\\
&+\frac{r_0^2(1-\m \Q)}{2\Q}\sinh^2l\sqrt{\Q(1-\m \Q)},
\ea\ee
where we made the replacement $\sqrt{ \L_\m}z_2=l\sqrt{\Q(1-\m \Q)}$. It follows that the HEE is 
 
\be\ba 
 S_{\text{EE}} 
=\frac{1}{4G}\cosh^{-1}&\left[\frac{\Q}{2r_0^2(1-\m \Q)}\sinh^2l\sqrt{\Q(1-\m \Q)}
+\cosh^2l\sqrt{\Q(1-\m \Q)}\right.\\
&+\left.\frac{r_0^2(1-\m \Q)}{2\Q}\sinh^2l\sqrt{\Q(1-\m \Q)} 
\right].
\ea\ee
For the $r_0\to \infty$ limit, note the definition of temperature~\eqref{temdef} and relation ${1}/{4G}={c}/{6}$, we arrive at 
\be \label{EEB1}
S_{\text{EE}}=\frac{c}{3}\log\left(\frac{\sqrt{\b^2+4\m\pi^2}+\b}{2\pi\epsilon}\sinh\left(\frac{\pi l}{\sqrt{\b^2+4\m\pi^2}}\right)\right),\quad \epsilon=\frac{1}{r_0}.
\ee
This coincides with~\eqref{HEEttbar} in the case of non-rotating BTZ black hole. We obtain the same holographic entanglement entropy formula by calculating the RT surface in the deformed BTZ black hole.
\section{Conclusion and discussion}
\label{sec:5}
The $T\bar{T}$ deformed CFT was proposed dual to the AdS$_3$ with a certain mixed boundary condition. The AdS$_3$ with mixed boundary condition or the $T\bar{T}$-deformed AdS$_3$ geometry can be obtained from the Ban\~ados geometry using the dynamical change of coordinates. In this paper, we studied the holographic entanglement entropy in the $T\bar{T}$-deformed AdS$_3$ under this situation. In terms of Chern-Simons form, we derived the holographic entanglement entropy formula using the Wilson line technique. For the zero temperature case, the entanglement entropy turned out unchanged under the $T\bar{T}$ deformation. For the finite temperature case, we calculated the Wilson line with ending points on the boundary of deformed AdS$_3$. After identifying the deformed temperature and length of interval on the boundary, we found the Wilson line lead to holographic entanglement entropy formula, which is closely related to the entanglement entropy in $T\bar{T}$-deformed CFTs. The same formula was also obtained by calculating the RT surface in the $T\bar{T}$-deformed BTZ black hole. The deformed entanglement entropy formula can reproduce the known perturbative results, which were obtained from both field theory and cutoff AdS$_3$ in other literature. We also showed that the entropic c-function is always positive and non–decreasing along the renormalization group flow towards the ultraviolet. For the non-perturbative region, our results show that the entanglement entropy behaves like entanglement entropy of CFT at zero temperature.
\par
Moreover, we also considered the two intervals entanglement entropy and found there still exist a certain phase transition between disconnected and connected phase. It turned out that the critical point for the phase transition depends on the deformation parameters. The critical point is sensitive to the deformation parameter for the high temperature region. But the critical point becomes independent of deformation parameter for the low temperature region. For a fixed temperature, the critical point tends to the zero temperature case at large deformation parameter, which is shown in Figure~\ref{p2}.  
\par
Finally, we want to point out that the holographic entanglement entropy formula was derived from the holographic study and the formula agrees with the perturbative result. However, we still need an exact calculation from $T\bar{T}$-deformed CFTs. In addition, since we found the entanglement entropy behaves like a free CFT, it would be interesting to study the $T\bar{T}$ deformation for large deformation  parameter following~\cite{Chakrabarti:2020pxr,Chakrabarti:2020dhv}.
\section*{Acknowledgements}
We are grateful to Song He for suggesting this topic. We would like to thank Yunfeng Jiang, Zhangcheng Liu, Hao Ouyang, Qiang Wen and Long Zhao for helpful discussions. This work is supported by the National Natural Science Foundation of China (No.12105113). 
\appendix
\section{Conventions}
\label{convention}
In this paper, we choose the following standard Lie algebra generators of $\mathfrak{sl}(2,\mathbb{R})$
\begin{align}
L_{-1}= \left[\begin{array}{ll}
0 & 1 \\
0 & 0
\end{array}\right],\quad 
L_{0} = \left[\begin{array}{cc}
\frac{1}{2} & 0 \\
0 & -\frac{1}{2}
\end{array}\right], 
\quad 
L_{1} & = \left[\begin{array}{cc}
0 & 0 \\
-1 & 0
\end{array}\right],
\end{align}
whose commutators simplify to
\begin{align} 
[L_a,L_b]=(a-b)L_{a+b},\quad a,b\in\{0,\pm 1\}.
\end{align}
The non-zero components of non-degenerate bilinear form are given by
\begin{align} 
\text{Tr}(L_0L_0)=\frac{1}{2},\quad \text{Tr}(L_{-1}L_1)=\text{Tr}(L_1L_{-1})=-1.
\end{align}
\par
We use the following representation of the $\mathfrak{sl}(2,\mathbb{R})$ Lie algebra, i.e. the highest-weight representation. The highest-weight state $|h\rangle$ satisfies
\begin{align}
L_{1}|h\rangle=0,\quad L_{0}|h\rangle=h|h\rangle.
\end{align}
There is an infinite tower of descendant states found by acting with the raising operator 
\begin{align}
|h,n\rangle=(L_{-1})^n|h\rangle.
\end{align}
These states form an irreducible, unitary, and infinite-dimensional representation of $\mathfrak{sl}(2,\mathbb{R})$.  The quadratic Casimir operator of the algebra is 
\begin{align}
C=2L_0^2-(L_1L_{-1}+L_{-1}L_1),
\end{align}
which commutes with all the elements of the algebra. The expectation value of Casimir operator on highest-weight state is 
\begin{align}
\mathcal{C}=\langle h|C|h\rangle=2h^2-2h.
\end{align}

\section{Wilson line defects}
\label{app:defect}
The Wilson line as a probe in the bulk will produce a back-reaction in the bulk. To solve for this back-reaction, we consider the total action 
\begin{align}
S=S_{CS}[A]-S_{CS}[\bar{A}]+B+S(U; A, \bar{A})_{C}.
\end{align}
where $B$ denotes the boundary term, the last term is the auxiliary action associated with the Wilson line. For different boundary conditions, there will be different boundary terms. In case of the $T\bar{T}$ deformation, the boundary term turns out to be 
\begin{align}
B=\frac{k}{4 \pi} \int_{\partial M} d^2x \frac{1}{\mu}\left(\sqrt{1-2 \mu\left(\text{Tr}(A_{\theta}{A}_{\theta})+\text{Tr}(\bar{A}_{\theta}\bar{A}_{\theta})\right)+\mu^{2}\left(\text{Tr}(A_{\theta}{A}_{\theta})-\text{Tr}(\bar{A}_{\theta}\bar{A}_{\theta})\right)^{2}}-1\right).
\end{align}
This boundary term leads to the $T\bar{T}$ deformed spectrum and can also help to reduce the gravitational action to $T\bar{T}$ deformed Alekseev-Shatashvili action on the boundary~\cite{He:2020hhm}. The boundary term does not contribute to the equation of motion, but the Wilson line term will contribute as a source for the equations of motion
\begin{align}
\label{defect-1}
\frac{k}{2\pi}F_{\mu\nu}=&\int ds\frac{dx^{\rho}}{ds}\varepsilon_{\mu\nu\rho}\delta^{(3)}(x-x(s))UPU^{-1},\\
\label{defect-2}
\frac{k}{2\pi}\bar{F}_{\mu\nu}=&-\int ds\frac{dx^{\rho}}{ds}\varepsilon_{\mu\nu\rho}\delta^{(3)}(x-x(s))P.
\end{align}
We can choose the Wilson line trajectory as a bulk geodesic, the corresponding Wilson line variables is
\begin{align}
r(s)=s,\quad U(s)=\textbf{1},\quad P(s)=\sqrt{2\mathcal{C}}L_0.
\end{align}
Contracting \eqref{defect-1} and \eqref{defect-2} with the tangent vector to the curve, we find the non-vanishing components of field strength $F,\bar{F}$ are tangent to the curve
\begin{align}
F_{\mu\nu}\frac{dx^{\mu}}{ds}=0,\\
\bar{F}_{\mu\nu}\frac{dx^{\mu}}{ds}=0.
\end{align}
Since we can always transform the AdS$_3$ solution into the Poincar\'e coordinate~\cite{Rooman:2000ei,Krasnov:2001cu}, we  just consider the Poincar\'e AdS$_3$. The solution is asymptotic AdS$_3$ in Poincar\'e coordinate
\begin{align}
A=&L(a_{\text{source}}+d)L^{-1},\quad L=e^{-\ln rL_0}e^{-zL_1},\\
\bar{A}=&R^{-1}(a_{\text{source}}+d)R,\quad R=e^{-\bar{z}L_{-1}}e^{-\ln rL_0},
\end{align}
where the coupling to the source is taken into account by
\begin{align}
a_{\text{source}}=\sqrt{\frac{\mathcal{C}}{2}}\frac{1}{k}\left(\frac{dz}{z}-\frac{d\bar{z}}{\bar{z}}\right)L_0.
\end{align}
With the help of the identities $\partial\frac{1}{\bar{z}}=\bar{\partial}\frac{1}{z}=\pi\delta^{(2)}(z,\bar{z})$, one can verify these connections satisfy the sourced equations of motion. The connections are flat except for where the Wilson line sources them. We can obtain the specific form of the gauge field
\begin{align}
A=&L_0\frac{dr}{r}+rL_{1}dz+\sqrt{\frac{\mathcal{C}}{2}}\frac{1}{k}\left(\frac{dz}{z}-\frac{d\bar{z}}{\bar{z}}\right)(L_0-rzL_1),\\
\bar{A}=&-L_0\frac{dr}{r}-rL_{-1}d\bar{z}+\sqrt{\frac{\mathcal{C}}{2}}\frac{1}{k}\left(\frac{dz}{z}-\frac{d\bar{z}}{\bar{z}}\right)(L_0-r\bar{z}L_{-1}).
\end{align}
This solution produces the metric
\begin{align}
ds^2=\frac{dr^2}{r^2}+\frac{r^2 \left(-\sqrt{2} \sqrt{\mathcal{C}} k \left(z d\bar{z}-\bar{z}dz\right)^2+\mathcal{C} \left(z d\bar{z}-\bar{z}dz\right)^2-2k^2 z\bar{z} dzd\bar{z}\right)}{2 k^2 z \bar{z}}.
\end{align}
Consider the map from plane to cylinder $(\tau,\vartheta)$
\begin{align}
z=e^{\tau+i\vartheta},\quad \bar{z}=e^{\tau-i\vartheta},
\end{align}
the metric becomes
\begin{align}
ds^2=&\frac{dr^2}{r^2}-r^2e^{2\tau} \left(d\tau^2+\frac{d\vartheta^2\left(\sqrt{2\mathcal{C}}-k\right)^2}{k^2}\right).
\end{align}
One can see this is precisely the metric for AdS$_3$ with a conical singularity surrounding the Wilson line. The boundary geometry with Wilson line back-reaction becomes the $n$-sheet cylinder if we set the defect angle to be $2\pi(1-\frac{1}{n})$. Then we can find the relation
\begin{align}
\frac{\sqrt{2\mathcal{C}}}{k}=(n-1)+O((n-1)^2).
\end{align}
Since the Wilson line action generates the $n$-sheet manifold, the partition function for $n$-sheet manifold can be written as 
\begin{align}
Z_{n}=\log W_{\mathcal{R}}(C)=-\sqrt{2\mathcal{C}}L(x_i,x_j),
\end{align}
therefore the entanglement entropy can be obtained
\begin{align}
S_{\text{EE}}=\lim_{n\to 1}\frac{1}{1-n}\log W_{\mathcal{R}}(C)=k L(x_i,x_j),
\end{align}
which coincides with the RT formula.
\par
The stress tensor corresponds to Poincar\'e AdS$_3$ vanishes, namely $\mathcal{L}=0$ in \eqref{banados}. For the BTZ black hole, the stress tensor is a constant. According to the transformation law of the stress-tensor, we can transform the stress tensor to a constant by using a conformal map. After rescaling the radial coordinate, the BTZ black hole becomes Poincar\'e AdS$_3$ geometry with different period of the time direction. For the deformed BTZ black hole, we can perform the following coordinate transformation to~\eqref{deformed ads}
\begin{align} 
w&=(1-\mu\mathcal{Q})\xi+\mathcal{Q}\bar{\xi},\\ 
\bar{w}&=(1-\mu\mathcal{\bar Q})\bar{\xi}+\mathcal{\bar{Q}}\xi,\\ 
r&=(1-\mu\mathcal{Q})(1-\mu\mathcal{\bar{Q}})\tilde{r}.
\end{align} 
so that the metric becomes the same as BTZ black hole
\begin{align} 
ds^2=\frac{d\tilde{r}^2}{\tilde{r}^2}+\tilde{r}^2\left(d\xi d\bar{\xi}+\frac{1}{\tilde{r}^2}\left(\mathcal{L}d\xi^2+\mathcal{\bar{L}}d\bar{\xi}^2\right)+\frac{\mathcal{L}\mathcal{\bar{L}}}{\tilde{r}^4}d\xi d\bar{\xi}\right).
\end{align}  
One should note that the temperature (the period of Euclidean time) is different from the original BTZ black hole. The above consideration for the holographic entanglement entropy still holds for BTZ black hole and deformed BTZ black hole.
\providecommand{\href}[2]{#2}\end{document}